\documentclass[a4paper,11pt]{article}
\pdfoutput=1 

\usepackage{jcappub} 

\usepackage{graphicx}
\usepackage{dcolumn,color}
\usepackage[dvipsnames]{xcolor}
\usepackage{bm,hyperref}
\usepackage[utf8]{inputenc}

\definecolor{armygreen}{rgb}{0.29, 0.33, 0.13}

\newcommand\myshade{80}
\colorlet{mylinkcolor}{ForestGreen}
\colorlet{mycitecolor}{Red}
\colorlet{myurlcolor}{violet}

\hypersetup{
  linkcolor  = mylinkcolor!\myshade!black,
  citecolor  = mycitecolor!\myshade!black,
  urlcolor   = myurlcolor!\myshade!black,
  colorlinks = true
}

\title{Discovery prospects of dwarf spheroidal galaxies for indirect dark matter searches}

\author[a,b,c]{Shin'ichiro Ando,}

\author[a,b]{Bradley J. Kavanagh,}

\author[c,a,b]{Oscar Macias,}

\author[b]{Tiago Alves,}

\author[b]{Siebren Broersen,}

\author[b]{Stijn Delnoij,}

\author[b]{Thomas Goldman,}

\author[b]{Jim Groefsema,}

\author[b]{Jorinde Kleverlaan,}

\author[b]{Ja{\"i}r Lenssen,}

\author[b]{Toon Muskens,}

\author[b]{Liam X. Palma Visser,}

\author[b]{Ebo Peerbooms,}

\author[b]{Bram van der Linden,}

\author[b]{and Sill Verberne}

\affiliation[a]{GRAPPA Institute, University of Amsterdam, 1098 XH
Amsterdam, The Netherlands}
\affiliation[b]{Institute for Theoretical Physics, 
University of Amsterdam, 1098 XH
Amsterdam, The Netherlands}
\affiliation[c]{Kavli Institute for the Physics and Mathematics of the
Universe (WPI), University of Tokyo, Kashiwa 277-8583, Japan}

\emailAdd{s.ando@uva.nl}
\emailAdd{b.j.kavanagh@uva.nl}
\emailAdd{o.a.maciasramirez@uva.nl}

\date{\today}

\abstract{
We study the prospects for the Large Synoptic Survey Telescope (LSST) to find new dwarf spheroidal galaxies in the Milky Way. Adopting models of Milky-Way halo substructure and
 phenomenological prescriptions connecting subhalos and
 satellite galaxies, we obtain surface brightness distributions of
 $V$-band magnitude that lead us to predict that LSST will discover tens to hundreds of dwarf spheroidal galaxies above its sensitivity. The soon-to-be-discovered dwarfs will be interesting targets for
 indirect searches of dark matter annihilation yields.
We forecast the distribution function of gamma-ray emission from dark
 matter annihilation in these objects, and discuss the detectability of these signals at both Fermi
 Large Area Telescope (LAT) and Cherenkov Telescope Array (CTA). By combining information from the predicted dwarf galaxies, we obtain an expected sensitivity to the annihilation cross section $\langle
 \sigma v \rangle$ of $10^{-26}$~cm$^3$~s$^{-1}$ (for dark matter particles of mass 10~GeV with Fermi-LAT) and $5\times 10^{-24}$~cm$^3$~s$^{-1}$ (for dark matter particles of mass 500~GeV with CTA).
We find that the current uncertainties in the mass measurement of the Milky-Way halo are relatively minor compared with the Poisson errors associated to drawing the most promising dwarfs from the underlying flux distribution.
}

\begin{document}

\maketitle
\flushbottom

\section{Introduction}
\label{sec:introduction}

Dark matter (DM) searches, especially for signatures of weakly interacting
massive particles (WIMPs), have been a strong scientific driver in
high-energy physics and astrophysics \cite{Bertone:2004pz,Bertone:2018xtm}.
WIMP annihilation and the resulting gamma-ray emission is expected from dense
celestial regions such as the Galactic center of our own Milky Way and
its satellite galaxies identified as dwarf spheroidals (dSphs) \cite{Strigari:2018utn}.
Although no discovery has yet been made, the indirect dark matter detection remains one of the most promising search methods \cite{Gaskins:2016cha}.
Current gamma-ray telescopes such as the Fermi Large Area
Telescope (LAT) \cite{Atwood:2009,Atwood:2013rka} have so far been able to probe down to the thermal WIMP annihilation cross section for dark matter particles below 100 GeV \cite{2012ApJ...761...91A,Ackermann:2015zua,Fermi-LAT:2016uux,Hoof:2018hyn}, while the upcoming Cherenkov Telescope Array (CTA) will probe unexplored regions of WIMP parameter space for TeV dark matter \cite{Carr:2015hta,Saturni:2019mxx}, where many interesting dark matter
candidates have been proposed (e.g. \cite{Cirelli:2015bda,Catalan:2015cna,Lefranc:2016fgn}).

Because of the lack of baryonic matter (interstellar gas and stars), dSphs are
believed to have the highest signal to noise ratio for DM searches in all the sky. 
In recent years, many new dSphs have been found with existing surveys such as the Sloan Digital Sky Survey (SDSS) \cite{2006AAS...20917805Z,2016MNRAS.463..712T}, the Dark Energy Survey (DES) \cite{Drlica-Wagner:2015ufc} and Gaia~\cite{2018arXiv181104082T}.
In the near future, the Large Synoptic Survey Telescope (LSST) \cite{lsstSRD,ivezic2008lsst,abell2009lsst,Drlica-Wagner:2019xan} is expected
to find even more dSphs.
In particular, if there are nearby dSphs discovered with LSST, we could use current
 or planned gamma-ray telescopes such as Fermi-LAT and CTA to further the sensitivities in searches for dark matter signatures. In this work, we study the prospects for the detection of new Milky Way dSphs with LSST and the consequences for indirect searches for WIMP DM.

Dwarf spheroidal galaxies are predicted to be formed in DM subhalos.
Even though the connection between subhalos and dSphs is not completely
understood, there are a number of phenomenological relations that are known to apply \cite{2013ApJ...765...22B,Garrison-Kimmel:2013eoa,Hargis:2014kaa,2018MNRAS.479.2853N,2018MNRAS.473.2060J}.
For example, it is generally expected that bright dSphs tend to form in heavy
subhalos with large circular velocity.
In this paper, we employ the semi-analytic models recently proposed in Refs.~\cite{Hiroshima:2018kfv,Ando:2019xlm} to generate a list of DM subhalos capable of hosting dSphs.  
These semi-analytic models have been shown to successfully
reproduce relevant subhalo properties obtained from DM-only simulations (such as the mass function) for a wide range of halo masses and redshifts. These models are hence expected to provide reliable and physically motivated predictions below the numerical resolution of current $N$-body simulations.
 
Our method for assigning a dSph to a subhalo follows the phenomenological prescription proposed in Ref.~\cite{Nadler:2018iux}. We first obtain the distributions of both $V$-band magnitude and surface brightness of the dSphs. Then, we explore the properties of dSphs that would be newly discovered with the LSST,  especially the expected DM self-annihilation signals. We also assess the uncertainties associated with the current Milky Way mass measurements and their impact on the sensitivity of DM searches. Lastly, we perform Monte Carlo simulations to draw samples of LSST dSphs, and use them to make sensitivity projections for both Fermi-LAT and CTA. Our analysis shows that with future surveys we can reach sensitivities to thermally-averaged cross sections on the order of $ \langle \sigma v\rangle \sim 10^{-26}$~cm$^3$~s$^{-1}$ for a WIMP mass of 10~GeV with Fermi-LAT and $\langle \sigma v\rangle \sim 10^{-24}$~cm$^3$~s$^{-1}$ for a WIMP mass of 500 GeV with CTA. Of course, each Monte Carlo simulation we perform corresponds to a \textit{single realisation} of a Milky Way-like galaxy. We therefore find that these limits are associated with realisation uncertainties of about one order of magnitude, originating from the Monte Carlo drawing of dSphs from their underlying distribution.

This paper is organized as follows. 
In Sec.~\ref{sec:Models and formulation}, we introduce models of subhalos and distribution functions of their various properties (density profiles, spatial distance, etc.). We also introduce the expected gamma-ray flux from dark matter annihilation and the subhalo--satellite connection. In Sec.~\ref{sec:Flux distributions of satellites}, we show the results of our semi-analytic model --- distribution of $V$-band magnitude, surface brightness, and the rate of DM annihilation. Then in Sec.~\ref{sec:Prospects for gamma-ray observation}, we discuss prospects of Fermi-LAT and CTA for constraining DM annihilation by using the LSST dSphs generated with Monte Carlo simulations. Finally, we conclude the paper in Sec.~\ref{sec:Discussion and conclusions}.

\section{Models and formulation}
\label{sec:Models and formulation}

\subsection{Subhalo models}
\label{sub:Subhalo models}

In Ref.~\cite{Hiroshima:2018kfv}, an analytic model is developed to account for the mass evolution of
DM subhalos as a result of tidal stripping during accretion (see also Refs.~\cite{Bartels:2015uba, Ando:2019xlm}). 
This analytic approach allows for calculations of the mass-loss rate
for arbitrary host mass and redshift and thus has an advantage over
numerical methods.
Assuming all tidal stripping occurs during a single orbit, the mass-loss
rate can be approximated by $\dot{m}=[m-m(r_{t})]/T_{r}$, where $T_{r}$ is the orbital period, $m$ the virial mass of the subhalo at accretion and $m(r_{t})$ the mass of the subhalo enclosed within the
truncation radius. 
A comparison between this simple Monte-Carlo approach and $N$-body simulations
for various redshifts and host masses shows agreement between both approaches~\cite{Hiroshima:2018kfv}.

We approximate the density profile of the subhalos with the
Navarro-Frenk-White (NFW) profile~\cite{Navarro:1996gj}:
\begin{equation}
 \rho(r) = \frac{\rho_s}{(r/r_s)(r/r_s+1)^2},
  \label{eq:NFW}
\end{equation}
for $r \le r_t$ and zero otherwise.
Each subhalo is then fully characterized by three parameters $\rho_s$,
$r_s$, and $r_t$.

We obtain these relevant parameters for the subhalos as follows.
Initially, we characterize each subhalo with the pre-infall mass $m_{200}$ at the accretion redshift $z_{a}$.
The scale radius $r_{s,a}$ and the scale density $\rho_{s,a}$ at accretion are then derived with the appropriate concentration-mass relation of the field halos~\cite{Correa:2015dva}.
Then we evolve the subhalo through tidal stripping. We obtain the mass of the subhalo $m_{0}$ at $z = 0$ by solving the differential equation for the subhalo mass, which is given per the host mass and redshift.
At the same time, we consider evolution of the density profiles characterized with $r_{s,0}$ and $\rho_{s,0}$ by following the empirical relations found in Ref.~\cite{2010MNRAS.406.1290P}.
We finally obtain the tidal truncation radius $r_{t,0}$ through mass conservation.
The list of the subhalos and their parameters is made such that both $\ln m_{200}$ and $z_{a}$ are scanned evenly.
However, those with smaller masses $m_{200}$ are much more abundant than heavier ones.
In order to correct for this, we assign each subhalo in the list effectively with a weight factor $w$, which is proportional to the number of accreted subhalos.
As the parameters after the tidal evolution ($m_0$, $r_{s,0}$, $\rho_{s,0}$, $r_t$) are deterministically obtained with the initial mass at accretion $m_{200}$ and the accretion redshift $z_a$ (as well as the concentration parameter), this weight $w$ is proportional to the probability distribution of $m_{200}$ and $z_a$, which is derived through the extended Press-Schechter theory~\cite{Yang:2011rf}.

We sample $\ln m_{200}$ and $z_a$ equally between $m_{200}=10^5M_\odot$ and 10\% of the host mass and between $z_a = 7$ and 0.1. In addition, subhalos satisfying the condition $r_{t,0}/r_{s,0}<0.77$ are considered to be tidally disrupted and are not included in the sample~\cite{Hiroshima:2018kfv}.

\subsection{Subhalo distribution}
\label{sub:Subhalo distribution}

The spatial distribution of subhalos is affected by a variety of processes -- driven by both baryons and dark matter within the host halo \cite{Calore:2016ogv,Kim:2017iwr}. Adiabatic contraction acts to increase the density in the center of the host galaxy \cite{Blumenthal:1985qy,Gnedin:2004cx} while tidal disruption acts to remove mass from the subhalos, a process which is more efficient in the dense central region of the host halo. In addition, the baryonic disk may disrupt subhalos which pass through it \cite{2010ApJ...709.1138D}. These effects mean that the distribution of subhalos throughout the galaxy will differ
from the distribution of the smooth host halo itself.  We follow Ref.~\cite{Calore:2016ogv}, which used hydrodynamical simulations to determine the spatial distribution  of subhalos. In this case, the number density of subhalos can be
approximated by the Einasto profile:
\begin{equation}
 n_{\rm sh}(r) \propto \exp
  \left[-\frac{2}{\alpha}
   \left(\frac{r}{r_{-2}}\right)^\alpha\right],
\end{equation}
where $\alpha = 2.2$ and $r_{-2} = 202$~kpc. We apply a further correction to this number density profile when calculating the radial distribution of \textit{observed} satellites, incorporating baryonic disruption and completeness effects~\cite{Kim:2017iwr}. \footnote{Figure~1 in Ref.~\cite{Kim:2017iwr} can be used to estimate the ratio of the cumulative radial distributions of satellites with and without baryonic disruption (light red and dotted black curves respectively). We correct our number density profile such that the ratio of cumulative distributions with and without correction matches that with and without baryonic disruption. See Ref.~\cite{Garrison-Kimmel:2017a} for more discussion about this effect.}




The procedure for converting the Galactocentric density of subhalos
$n_{\rm sh}(r)$ to the average heliocentric distance distribution
$P_{\rm sh}(D)$ is as follows.
First, a formula is found to express the distance from the Galactic
center to the subhalo as a function of the angle $\psi$ between the direction to the subhalo and
that to the Galactic centre.
This result is substituted into the distribution of subhalos
$n_{\rm sh}(r(D,\psi))$ and by averaging over a sphere of radius $D$ we
find the average number density
\begin{equation}
\overline{n}_{\rm sh}(D)=\frac{1}{2}\int_{-1}^{1}n_{\rm sh}(r(D,\psi))
 \,\mathrm{d}\cos\psi.
\label{n_sh_avg}
\end{equation}
To find the probability of finding a subhalo at a certain distance $D \rightarrow D + \mathrm{d}D$, it
is necessary to multiply by the volume element 4$\pi D^2\,\mathrm{d}D$. 
The resulting function for the probability density of finding a subhalo
at a certain heliocentric distance is then $P_{\rm sh}(D) \propto D^{2}n_{\rm
sh}(D)$.

\subsection{Astrophysical \textit{J}-factor}
\label{sub:Astrophysical J factor}

The gamma-ray flux due to dark matter annihilation from a dSph at
distance $D$ is given by
\begin{equation}
 F_\gamma(E) = \frac{\langle\sigma
  v\rangle}{2m_\chi^2}\frac{\mathrm{d}N_{\gamma,{\rm ann}}}{\mathrm{d}E}
  \frac{1}{4\pi}\int_{\Delta\Omega} \mathrm{d}\Omega \int \mathrm{d}\ell
  \rho^2_{\chi}(r(\ell,\Omega)),
\label{eq:flux}
\end{equation}
where $r$ is the radial coordinate of the dSph, $r^2(\ell,\psi)=\ell^{2}+{D}^{2}-2D \ell \cos\psi$, $\langle \sigma v\rangle$ is the annihilation cross section times
relative velocity, $m_\chi$ is the mass of the dark matter particle
$\chi$, $\mathrm{d}N_{\gamma,{\rm ann}}/\mathrm{d}E$ is the gamma-ray spectrum per
annihilation, $\Delta\Omega$ is the solid angle over which we compute
the gamma-ray flux, and $\ell$ is a line-of-sight parameter.
The latter factor in Eq.~\eqref{eq:flux} is often referred to as the astrophysical $J$-factor:
\begin{equation}
 J = \int_{\Delta\Omega} \mathrm{d}\Omega \int \mathrm{d}\ell \rho_\chi^2(r(\ell,\Omega)).
\end{equation}

Reference~\cite{Evans:2016xwx} provides an accurate expression for $J$ as a function of the properties of a DM halo.
This expression can be obtained by assuming that the distance to the
dSph is large enough such that the integration variables can be
appproximated as
\begin{equation}
\mathrm{d}\Omega \,\mathrm{d}\ell\approx \frac{1}{D^2}2\pi R\, \mathrm{d}R \,\mathrm{d}z,
\label{intvar}
\end{equation}
where $R$ is the polar coordinate in the plane of the sky and $z$ the
line of sight. 
It is useful to define an auxiliary function $X(s)$ which is given by
\begin{align}
X(s)=\begin{cases}\frac{1}{\sqrt{1-s^2}}\text{arcsech}(s), &0\leq s\leq1 \\ 
\frac{1}{\sqrt{s^2-1}}\text{arcsec}(s), &s\geq1 \end{cases}
\label{X}
\end{align}
with $X(1)=1$ to make sure that the function is continuous. 
Combining Eqs.~(\ref{intvar}) and (\ref{X}) and integrating from $z=-\infty$ to
$\infty$ and from $R=0$ to $D\theta$ leads to
\begin{eqnarray}
J(\rho_s, r_s, r_t, D) &=&\frac{\pi \rho_{s}^2r_{s}^3}{3D^2\Delta^4}\left[2y(7y-4y^3+3\pi
					       \Delta^4) +6(2\Delta^6-2\Delta^2-y^4)X(y)\right]\,,
\label{Jacc}
\end{eqnarray}
where $\theta$ is the integration angle, $y=D\theta/r_{s}$ and
$\Delta^2=1-y^2$. We extend the integral out to the angular size of the halo, $\theta = \arctan(r_t/D)$, or $\theta = 0.5^\circ$ if the angular size is larger than $0.5^\circ$. We note that these expressions are valid for subhalos with NFW density profiles; see Ref.~\cite{Hiroshima:2019wvj} for a discussion of the impact of density profile uncertainties on projected constraints.

We use Eq.~\eqref{Jacc} to determine the $J$-factor for a given subhalo.
By combining the joint probability distribution function (PDF) of $\rho_s$, $r_s$, and
$r_t$ (Sec.~\ref{sub:Subhalo models}) and an independent distribution for the heliocentric
distance $D$ (Sec.~\ref{sub:Subhalo distribution}), we can then compute
the $J$-factor distribution in a Milky Way-like galaxy:
\begin{equation}
P(J) = \int \mathrm{d}\rho_s \int \mathrm{d}r_s \int \mathrm{d}r_t \int  \mathrm{d}D \, \Big[P(\rho_s, r_s, r_t) P(D) \delta\left(J - J(\rho_s, r_s, r_t, D) \right)\Big]\,.
\end{equation}

\subsection{Subhalo-satellite connection}
\label{sub:Subhalo-satellite connection}

Satellite galaxies can form in large dark matter subhalos.
A subhalo can be seen as a strong gravitational well that attracts
baryonic matter. 
With enough matter attraction, star formation can be possible and a
dSph can be formed. 
The luminosity of such a satellite is believed to be well correlated
with the peak maximum circular velocity of the subhalo, $V_{\rm peak}$,
or alternatively the mass at accretion. 
The maximum circular velocity $V_{\rm max}$ and the radius $r_{\rm max}$ at
which this maximum velocity is reached are obtained with $\rho_s$ and
$r_s$ via \cite{Hiroshima:2018kfv}
\begin{eqnarray}
    r_{\rm max} &=& 2.163 r_s,
    \label{rmax}\\
    V_{\rm max} &=& \left( \frac{4 \pi G \rho_s}{4.625}\right)^{1/2}
     r_s.
    \label{vmax}
\end{eqnarray}
Both $\rho_s$ and $r_s$ evolve after the subhalo has accreted onto the host,
and in our model, $V_{\rm max}$ is largest at the moment when the
subhalo has just accreted:
\begin{equation}
    V_{\rm peak} = \left( \frac{4 \pi G \rho_{s,
		    a}}{4.625}\right)^{1/2} r_{s,a}.
\end{equation}

Recently, Ref.~\cite{Nadler:2018iux} presented a model for mapping
dark matter subhalo properties to those of corresponding satellites.
Methods for converting the subhalo property $V_{\rm peak}$ to the
half-light radius of a satellite, $r_{1/2}$, and V-band magnitude,
$M_V$, are described below.
We use the relation
\begin{equation}
    r_{1/2} \equiv \mathcal{A} \left( \frac{c}{10} \right)^{\gamma}
     R_{\rm vir} ,
    \label{halflightradius}
\end{equation}
to determine the half light radius at accretion, where $c$ denotes the
subhalo concentration $R_{\rm vir} / r_{s, {\rm acc}}$ at accretion,
$\mathcal{A} = 0.02$ and $\gamma = -0.7$.
From this, Ref.~\cite{Nadler:2018iux} introduced the relation
\begin{equation}
    r'_{1/2} = r_{1/2}\left( \frac{V_{\rm max}}{V_{\rm peak}} \right)^\beta,
    \label{halfligtradius_at_z=0}
\end{equation}
to determine the half-light radius at $z = 0$. 
Here $V_{\rm max}$ denotes the peak circular velocity at $z=0$. 
In this equation, $\beta \geq 0$ is a model parameter that denotes the
change in radius caused by tidal stripping; in our model we use a fixed
value of $\beta = 1$. 
Using abundance matching, Ref.~\cite{Nadler:2018iux} found a direct
relation between $V_\mathrm{peak}$ and $M_V$. 
Further, we use the relation
\begin{equation}
    \label{surfacebrightness}
    \mu_V = M_V + 36.57 + 2.5\log{[2 \pi (r'_{1/2} / 1 \: \text{kpc} )^2 ]},
\end{equation}
for calculating the surface brightness.
Using Eq.~(\ref{surfacebrightness}), a cumulative distribution of surface
brightness is obtained, which is then used to determine the number of
observable satellites for a telescope with a specific sensitivity.

Whether a subhalo forms a satellite inside or not is subject to a
formation threshold. 
It is assumed that the subhalo must have a minimum peak velocity for
galaxy formation. 
Following Ref.~\cite{Nadler:2018iux}, we use a galaxy formation threshold
$V_{\rm peak}$ in the range between 14 and 22 km~s$^{-1}$.
These models for connecting subhalo parameters to satellites are
consistent with the observed Milky Way satellite population.

\section{Flux distributions of satellites}
\label{sec:Flux distributions of satellites}

\subsection{Distribution of surface brightness of satellites}
\label{sec:surface_brightness}

\begin{figure}[t]
\begin{center}
\includegraphics[width=8.5cm]{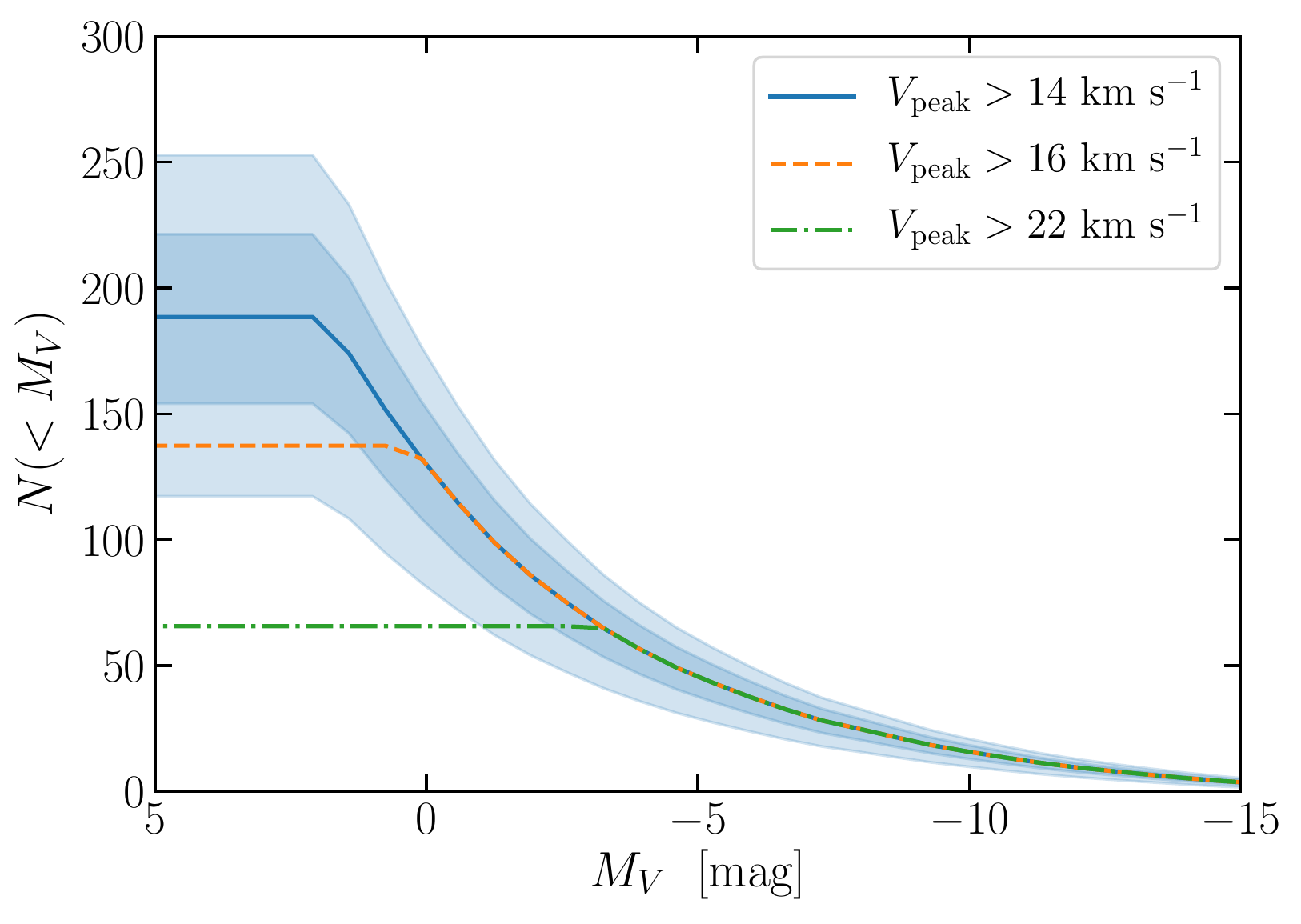}
\caption{Cumulative number of satellite galaxies over the whole sky below
 a given absolute magnitude $M_V$ for different thresholds on $V_{\rm
 peak}$. The bands attached to the curve for $V_{\rm
 peak}>14~\mathrm{km~s^{-1}}$ represent 1$\sigma$ and 2$\sigma$
 uncertainties related to the Milky-Way halo mass measurement $M_{200} =
 1.3\pm 0.3\times 10^{12} M_\odot$.}
\label{mv_distribution}
\end{center}
\end{figure}

We first discuss the cumulative distribution of the absolute V-band magnitude $M_V$,
which is shown in Fig.~\ref{mv_distribution} for the case of $M_{200} =
1.3\times 10^{12}M_\odot$.
For $V_{\rm peak}$ thresholds of 14, 16, and 22~km~s$^{-1}$, we expect a
total number of the satellites of $\sim$190, $\sim$140 and $\sim$70,
respectively.
The bands attached to the curve for $V_{\rm peak}>
14~\mathrm{km~s^{-1}}$ accommodate 1$\sigma$ and 2$\sigma$ uncertainties
in relation to the mass measurement of the Milky-Way halo,
$\sigma_{M_{200}} \approx 0.3 \times
10^{12}M_\odot$~\cite{2019AA...621A..56P} (see also
Refs.~\cite{2018ApJ...863...89S, 2018arXiv180411348W,
2018AA...616L...9M}).\footnote{We, however, do not accommodate
uncertainties arising from different halo accretion history, which
earlier work estimates to be around the tens of percent
level~\cite{Jiang:2016yts}.}
This result, especially for the former two cases, is consistent with
$134\pm 44$ with $M_V < 0$ found by Ref.~\cite{Nadler:2018iux} as well
as results of other earlier work~\cite{Tollerud:2008ze, Koposov:2007ni,
2018MNRAS.479.2853N}.

\begin{figure}[t]
\begin{center}
\includegraphics[width=8.5cm]{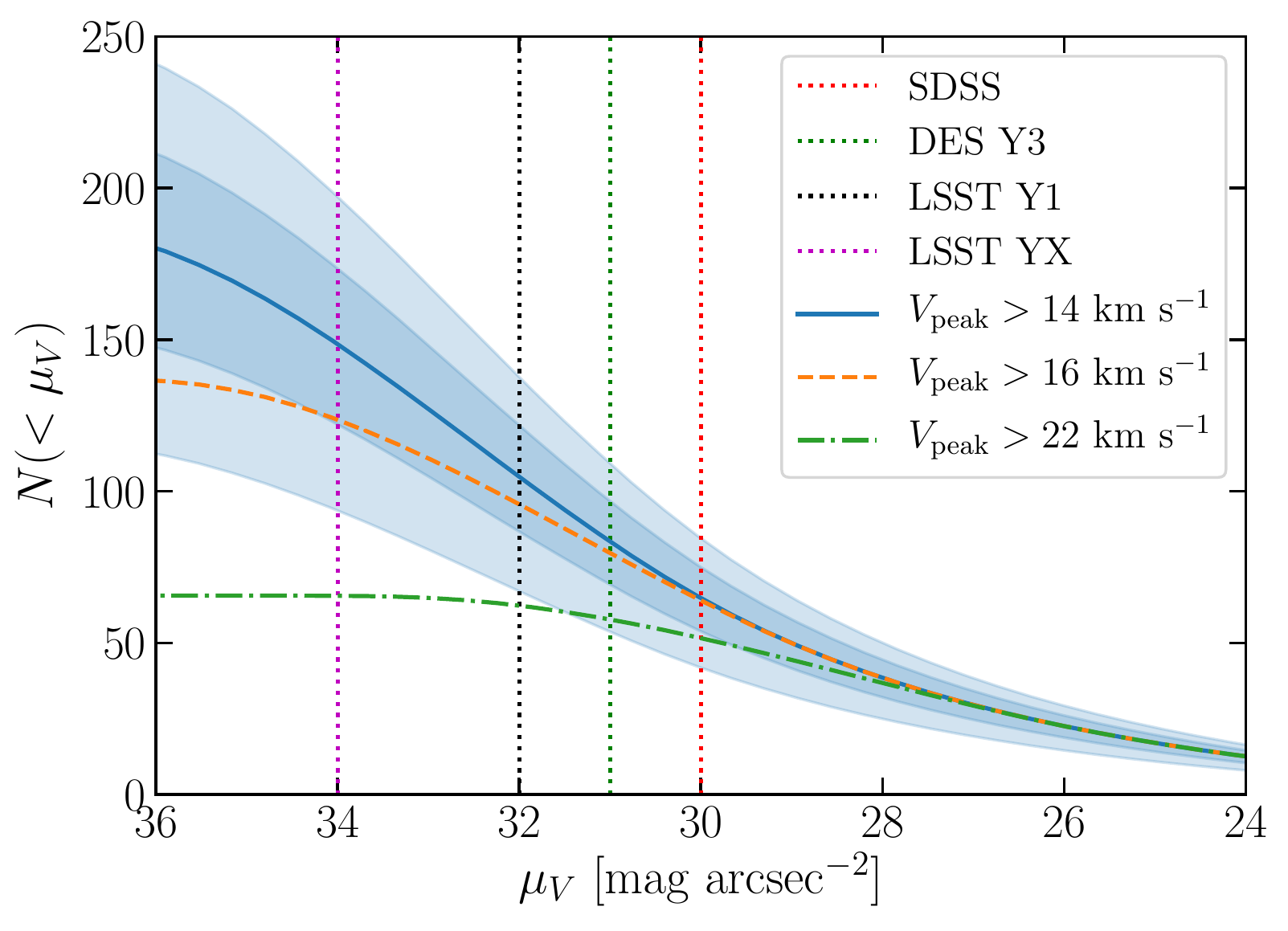}
\caption{Cumulative number of satellite galaxies over the whole sky below 
 given surface brightness $\mu_V$, for different thresholds on $V_{\rm
 peak}$. The vertical dotted lines are the different surface brightness
 limits corresponding to SDSS, DES, and LSST. The bands attached to the
 curve for $\mu_V<34$ represents the Milky Way mass uncertainties as in
 Fig.~\ref{mv_distribution}.}
\label{surface_brightness_distribution}
\end{center}
\end{figure}

Since galaxies are an extended source, the surface brightness $\mu_V$
[Eqs.~(\ref{halfligtradius_at_z=0}) and (\ref{surfacebrightness})] is
a more relevant quantity for discussing detection prospects, which we
show in Fig.~\ref{surface_brightness_distribution}.
For the surveys SDSS, DES and LSST year 1, following
Ref.~\cite{Nadler:2018iux}, we adopt a surface brightness threshold of
30, 31, and 32~mag~arcsec$^{-2}$, respectively.
For an extrapolation to further years of LSST, we looked at historical
data from SDSS which has been releasing data since 2003. 
For every data release we looked at the faintest detected source in
terms of Petrosian magnitude. 
In data release 1~\cite{Abazajian:2003jy}, the faintest detected
source was about 2 magnitudes brighter than in the data release
15~\cite{2019ApJS..240...23A}, for which most of this improvement was
made between the first and the second release.
We will adopt this improvement of 2 magnitudes as an indication to how
much LSST can be expected to improve after its first year of
operation.
The resulting limiting surface brightnesses of these surveys are shown
as vertical dotted lines in Fig.~\ref{surface_brightness_distribution}.

These results are again in good agreement with
Ref.~\cite{Nadler:2018iux}, which predicted $92\pm 29$ satellites with
$M_V < -1.5$~mag (that corresponds roughly to $V_{\rm peak} >
18$~km~s$^{-1}$) for surface brightness limit of 34~mag~arcsec$^{-2}$.
We predict a factor of $\sim$2 improvement of the total number of
satellites (over the whole sky) detected with the LSST over that with the
DES.
The difference is more pronounced for smaller values of threshold
$V_{\rm peak}$, as in that case, there will be many more faint satellites.

The main survey of LSST will cover about $\sim$18000~deg$^2$ of the
sky~\cite{footprint}, compared to around 5000~deg$^2$ for DES \cite{DESfootprint}: taking this effect into account, the difference between the two will be even larger. We finally note  that LSST's footprint covers almost the entire Southern sky. In contrast, SDSS covered predominantly the Northern sky \cite{2011ApJS..193...29A}. 
In all the cases we have investigated, it is expected that the LSST over
many years of operation will discover most of the satellites in the
Milky-Way halo within its footprint, the majority of which are as-yet undiscovered.


\subsection{Distribution of \textit{J} factor}
\label{sec:Jfactor_distribution}

\begin{figure}[t]
\centering
\includegraphics[width=8.5cm]{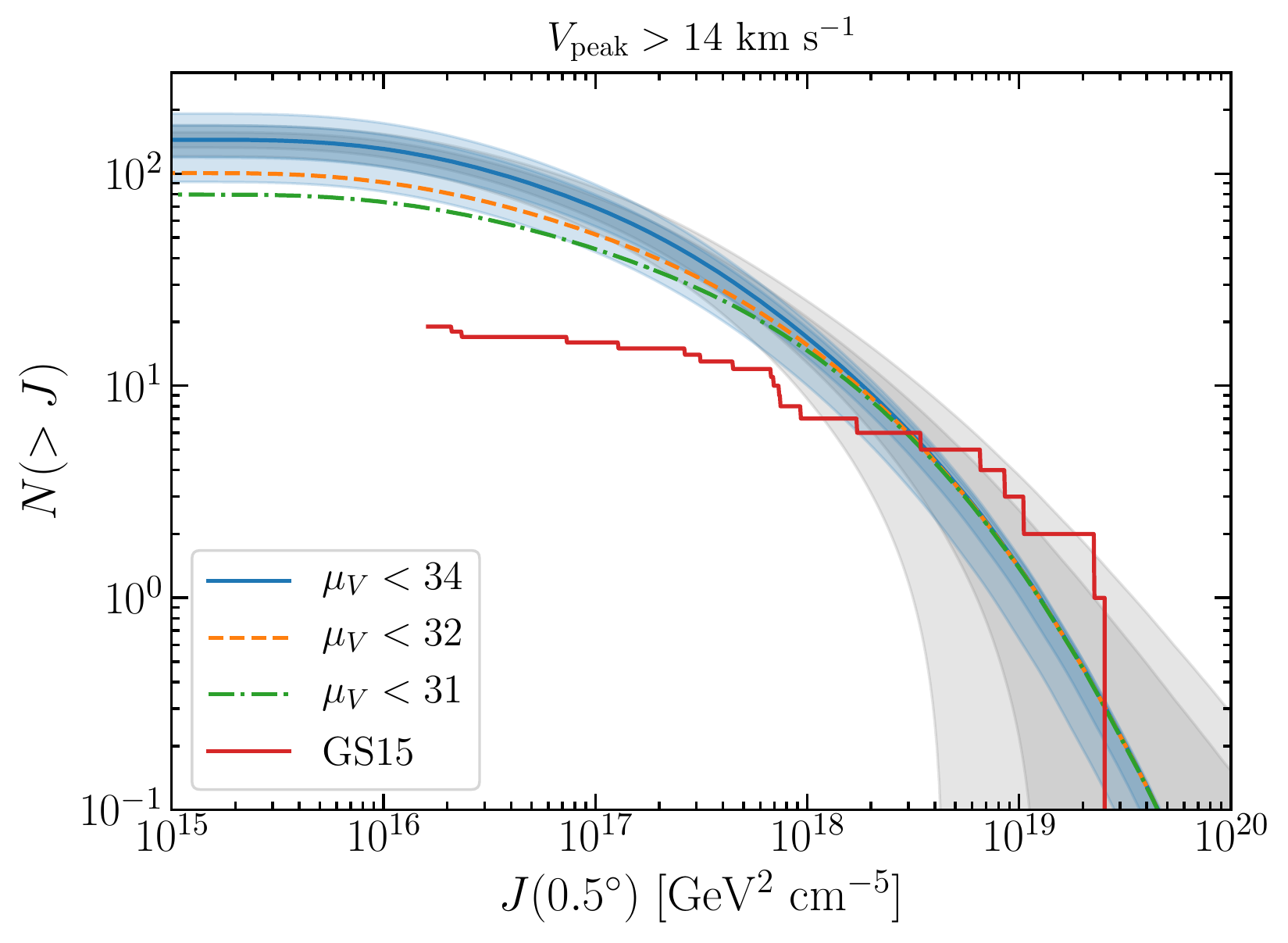}
\caption{Cumulative number of all-sky satellite galaxies (with $V_{\rm
 peak} > 14$~km~s$^{-1}$) above given values of $J$ within 0.5$^\circ$,
 for different values of threshold $\mu_V$. The blue error bands show
 Galaxy mass uncertainties as in Fig.~\ref{mv_distribution}. The gray
 bands are 1$\sigma$ and 2$\sigma$ Poisson errors, where $\sigma_N =
 \sqrt{N(>J)}$. The distribution of median $J$ values of the known dSphs
 according to Ref.~\cite{Geringer-Sameth:2014yza} (GS15) is shown as a
 solid histogram for comparison; see however the text for possible caveats.}
\label{fig:J_dist_muV}
\end{figure}

Next, we calculate the distribution of the astrophysical $J$-factor for all dSphs in the sky.
Figure~\ref{fig:J_dist_muV} shows the cumulative distribution $N(>J)$ of
dSphs with $V_{\rm peak}>14$~km~s$^{-1}$ that would be detected by
several galaxy surveys.
We show uncertainties related to the mass measurement of the Milky-Way
halo of $M_{200} = (1.3\pm 0.3)\times 10^{12}M_\odot$ as blue bands
attached to the curve for $\mu_V<34$.
We also show, as the gray bands, the Poisson uncertainties (both
1$\sigma$ and 2$\sigma$ levels), where $\sigma_N = \sqrt{N(>J)}$, which
is associated to the probability of drawing a certain number of dSphs from the
underlying distribution.
The figure shows that the uncertainty related to the Milky-Way halo mass
measurement is subdominant compared with the Poisson errors, especially
for dSphs with $J$ greater than $\sim$10$^{18}$~GeV$^{2}$~cm$^{-5}$,
which are the most interesting for indirect dark matter searches.
It also shows that compared with the sensitivity of the current
generation of the surveys such as DES, the LSST does not provide much improvement for finding dSphs with large $J$.
However, since the LSST footprint is expected to be much larger than
that of DES or any other existing surveys of relevance, it should still be able to discover many new dSphs with large $J$ values.

\begin{figure}[t]
\centering
\includegraphics[width=8.5cm]{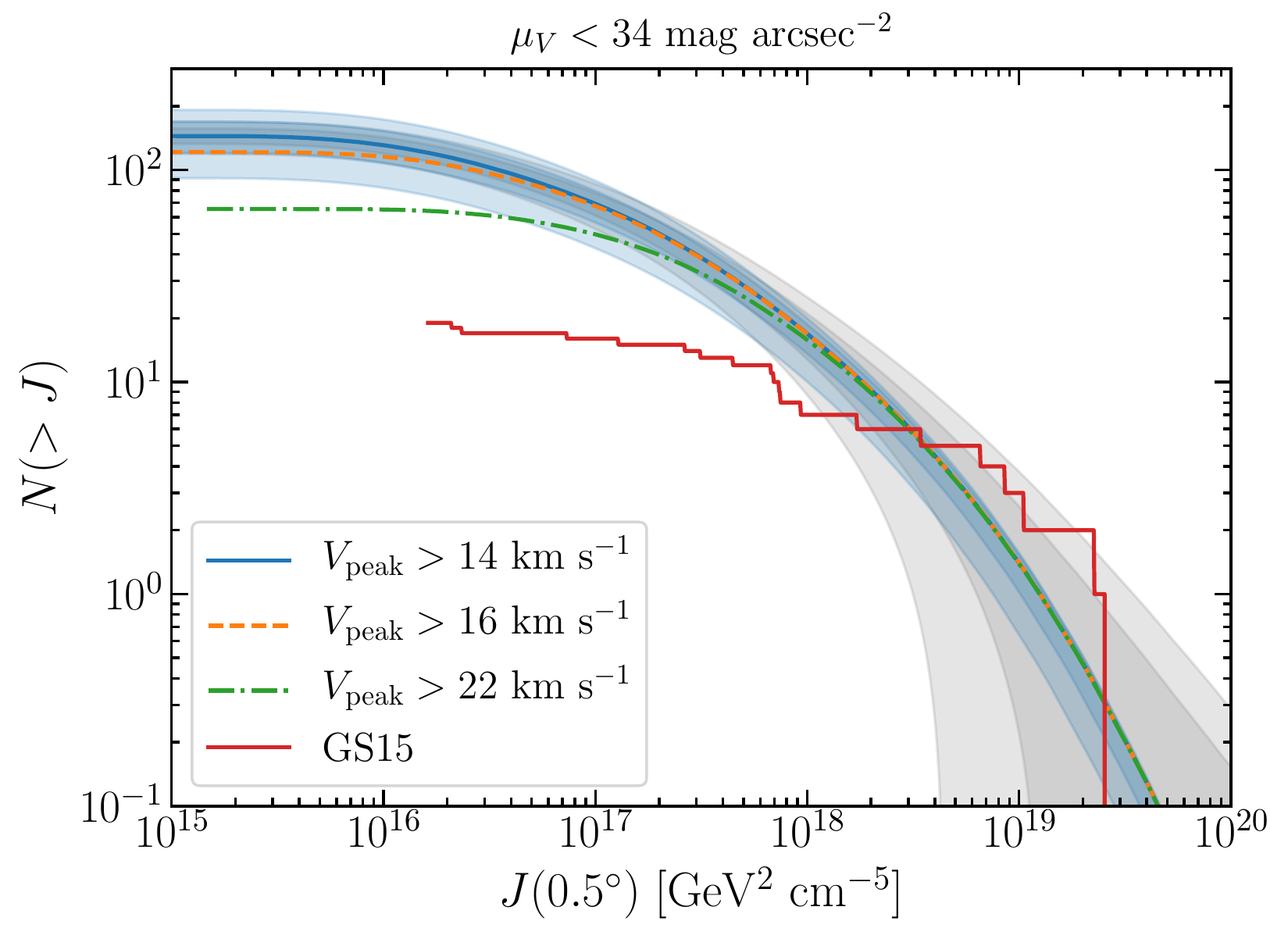}
\caption{The same as Fig.~\ref{fig:J_dist_muV} but for $\mu_V<34$ and
 various values of $V_{\rm peak}$.}
\label{fig:J_dist_Vpeak}
\end{figure}

Figure~\ref{fig:J_dist_Vpeak} then shows the dependence on $V_{\rm
peak}$ of the $J$ distribution.
We find that $V_{\rm peak}$ has only a small impact on $J$ distribution
especially for dSphs in interesting regimes with $J>
10^{18}$~GeV$^2$~cm$^{-5}$.
Therefore, the precise conditions for forming satellites in subhalos are not 
crucial for determining the prospects for indirect dark matter searches with dSphs
discovered by LSST.

In both Figs.~\ref{fig:J_dist_muV} and \ref{fig:J_dist_Vpeak}, we show
cumulative distributions of $J$ for the known
dSphs~\cite{Geringer-Sameth:2014yza}.
It is higher than the theoretical prediction at very large $J$ values
at the 2$\sigma$ level, which might simply be because of a statistical upward fluctuation.
However, we also note a few possible caveats.
First, we show only median values of $J$ in the figures, while errors in $J$ can be large, typically ranging from the tens of percent level to a factor of a few.
Next, the $J$-factors in Ref.~\cite{Geringer-Sameth:2014yza} might be overestimated; 
see Ref.~\cite{Hayashi:2016kcy} for axi-symmetric modeling that indicates smaller values of $J$-factors.
Furthermore, adopting more informative priors for the density profiles of the satellites likely shifts the $J$-factor estimates of these existing dSphs toward smaller values possibly by a factor of a few \cite{Ando_inprep}. In any case, it is expected that the LSST should find many more new dSphs with $J >
10^{18}$~GeV$^2$~cm$^{-5}$, enlarging the current sample by a factor of 2 or 3.

\section{Prospects for gamma-ray observation}
\label{sec:Prospects for gamma-ray observation}

\subsection{Generating mock dSphs}

Now that we have obtained the distribution of both $\mu_V$ and $J$, which are
relevant for LSST and gamma-ray telescopes, respectively, we now assess
the detectability of the gamma-ray emission from the LSST dSphs with
Fermi-LAT and CTA.

To illustrate what the expected sensitivities would be, we generate both
$J$ and $\mu_V$ for dSphs with Monte Carlo simulations drawn from each
respective distribution.
Here, since the errors associated with the Milky-Way mass measurement is
subdominant compared with the Poisson errors, we first randomly generate the 
number of dSphs ($N_{\rm sat}$) above threshold of $J_{\rm th} =
10^{15}$~GeV$^2$~cm$^{-5}$, drawing from the Poisson distribution with
a mean of $N(>J_{\rm th})$ in the case of $\mu_V < 34$ and $V_{\rm
peak}> 14$~km~s$^{-1}$.
Then, Monte Carlo simulations are used to assign each of these
$N_{\rm sat}$ dSphs a random value of $J$ following its distribution as
well as a sky location following the surface density along an arbitrary
direction.
These sky locations are checked to determine if they fall within the LSST footprint; if
not, these dSphs are removed from the subsequent projection analysis.

\subsection{Gamma-ray telescopes}

To estimate the sensitivity to dark matter annihilation of both
Fermi-LAT and CTA with newly discovered LSST dSphs, we utilize the
software package {\sf swordfish}, a python tool used to study the
information yield of event-counting experiments~\cite{edwards2018fresh,
edwards2017swordfish}. This allows us to perform rapid projections in scenarios
where large numbers of dSphs are detected.
To determine the expected upper limit, it is necessary to provide
information regarding the signal and background spectra as well as the
fractional uncertainty associated with the latter.

\subsubsection{Fermi-LAT}

Most of the gamma-ray photons detected by the Fermi-LAT correspond to the Galactic diffuse emission (GDE) generated by the interaction of energetic cosmic-ray particles with interstellar gas, ambient photons and magnetic fields~\cite{Atwood:2009}. Searches for a putative dark matter signal in the gamma-ray sky strongly depend on the accuracy and reliability of our model for the GDE. 

The Fermi team has developed and made public a sophisticated model for the GDE~\cite{Acero:2016qlg} that is suitable for our study. This was constructed by assuming that the measured diffuse gamma-ray intensity can be modeled as a linear superposition of interstellar gas column density maps describing the hadronic and bremsstrahlung components as well as an inverse Compton emission template created with the cosmic ray propagation code \textsf{GALPROP}~\cite{Galprop}. Furthermore, the Fermi GDE model contains some empirical templates to correct for regions where extended positive gamma-ray residuals have been detected. One notable example is the inclusion of a spatial template for the Fermi bubbles~\cite{Su:etal2010,FermiLat:Bubbles}, which are giant non-thermal structures extending up to $\sim$7~kpc above and below the Galactic plane. Due to their large angular size, these could be an important source of uncertainties for searches of dark matter emission from future LSST dSphs. 

In the present work we estimate the background and foreground gamma rays relevant to this analysis by using
the most recent GDE model \textsf{gll\_iem\_v07.fits}\footnote{The interested reader is referred to the cicerone for more details: \url{https://fermi.gsfc.nasa.gov/ssc/data/access/lat/BackgroundModels.html}.} released by the Fermi team and the corresponding isotropic emission model \textsf{iso\_P8R3\_SOURCE\_V2\_v1.txt} accounting for unresolved gamma rays (thought to be mostly of extragalactic origin). In our procedure, we conservatively apply a 20\% spectral normalization uncertainty to the total isotropic background, as well as an independent 20\% normalization uncertainty for the GDE at the sky position of each sampled dwarf.

The dark matter annihilation spectrum into gamma rays resulting from the decay of final state bottom quarks is determined with the package \textsf{PPPC 4 DM ID}~\cite{Cirelli:2011, Ciafaloni:2011}. This is then evaluated in Eq.~(\ref{eq:flux}) to obtain the expected dark matter flux emitted from a given dSph. As the next step, we compute the expected DM signal and background/foreground photon counts by convolving the resulting flux maps with the LAT instrument response function~\cite{Atwood:2009}. We assume an effective exposure of 2 years for all dSphs (corresponding to 10 years of observation with roughly $20\%$ sky coverage).  Lastly, both signal and background photon counts are fed to \textsf{swordfish} in order to compute the Fermi-LAT sensitivity to dark matter self-annihilations from future LSST dSphs. 




\subsubsection{CTA}


We explore the sensitivity of the forthcoming Cherenkov Telescope Array (CTA)~\cite{Consortium:2010bc} in a similar manner as for the Fermi-LAT.
However, in this case, we are required to construct a model for the GDE at the energy range of interest with the use of \textsf{GALPROP v54}~\cite{Galprop}. In particular, we use the results of the propagation parameter scan carried out by Ref.~\cite{Ackermann:2012} to choose a propagation parameter setup that reproduces well gamma-ray data in the 0.1--500~GeV energy range. The parameters assumed in our \textsf{GALPROP} simulation are: a cosmic ray source distribution as given by the supernova remnants, cosmic ray halo height of 4~kpc, an atomic hydrogen spin temperature of 150~K and a $E(B -V )$ magnitude cut of 5~mag. Since the physical mechanisms giving rise to the photon emission are thought to be the same regardless of the energy band of interest, the GDE simulated in this way is expected to provide an approximate picture of the gamma-ray sky in the CTA energy range that is better motivated than a simple extrapolation to higher energies of the Fermi GDE model.


We also include a background coming from the misidentification of charged cosmic rays. We follow the procedure of Ref.~\cite{Silverwood:2014yza}, assuming that cosmic ray electrons cannot be distinguished from gamma rays and that 1\% of cosmic-ray protons will be reconstructed as gamma rays (though with a factor of 3 smaller energy). The latter contribution becomes dominant above energies of around 2 TeV. We assume a 20\% normalization uncertainty for this cosmic-ray misidentification background as well as independent 20\% uncertainties on the GDE at the sky position of each dSph, as in the case of Fermi-LAT.

Since the effective area of the CTA is also energy dependent, a fit has
been made in the energy range of interest, which corresponds to the
interpolation of the results given in Ref.~\cite{bernlohr2013monte}.
We apply 50 hours of observation time for each dSph, which is in
agreement with the approach taken in Ref.~\cite{Lefranc:2016fgn}.

\subsection{Projected upper limits on the annihilation cross section}

\begin{figure}[t]
\begin{center}
\includegraphics[width=8.5cm]{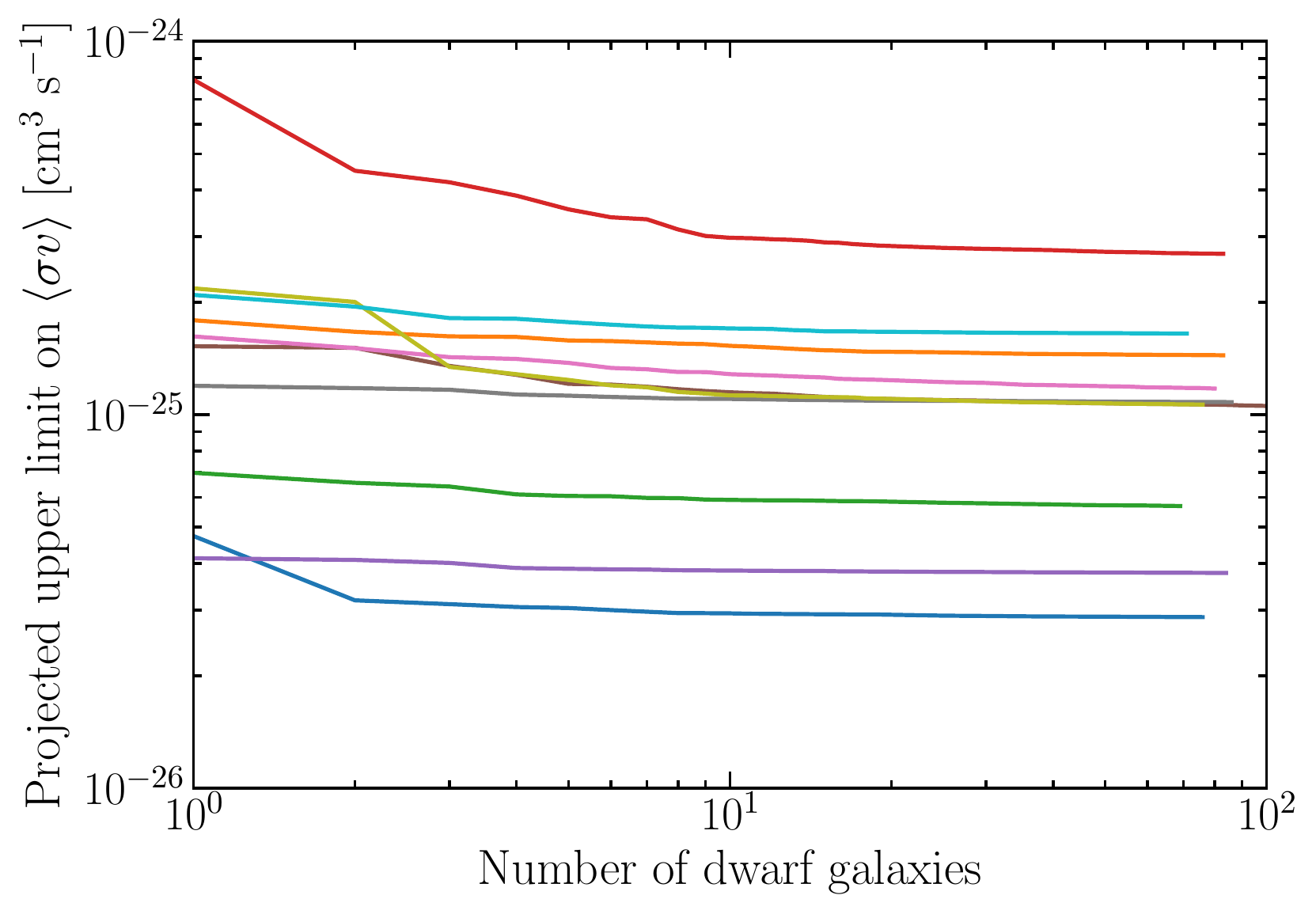}
\caption{Projected cross section upper limits at 95\% confidence level as a
 function of the number of dSphs used, sorted in order from the highest
 to lowest $J$ values. The results of ten random Monte Carlo sampling are shown. Cross section values are calculated for the
 Fermi-LAT, assuming annihilation into $b\overline{b}$ and a dark matter mass of 100 GeV.}
\label{fig:CSJFac}
\end{center}
\end{figure}

\begin{figure}[htbp]
\centering
\includegraphics[width=8.5cm]{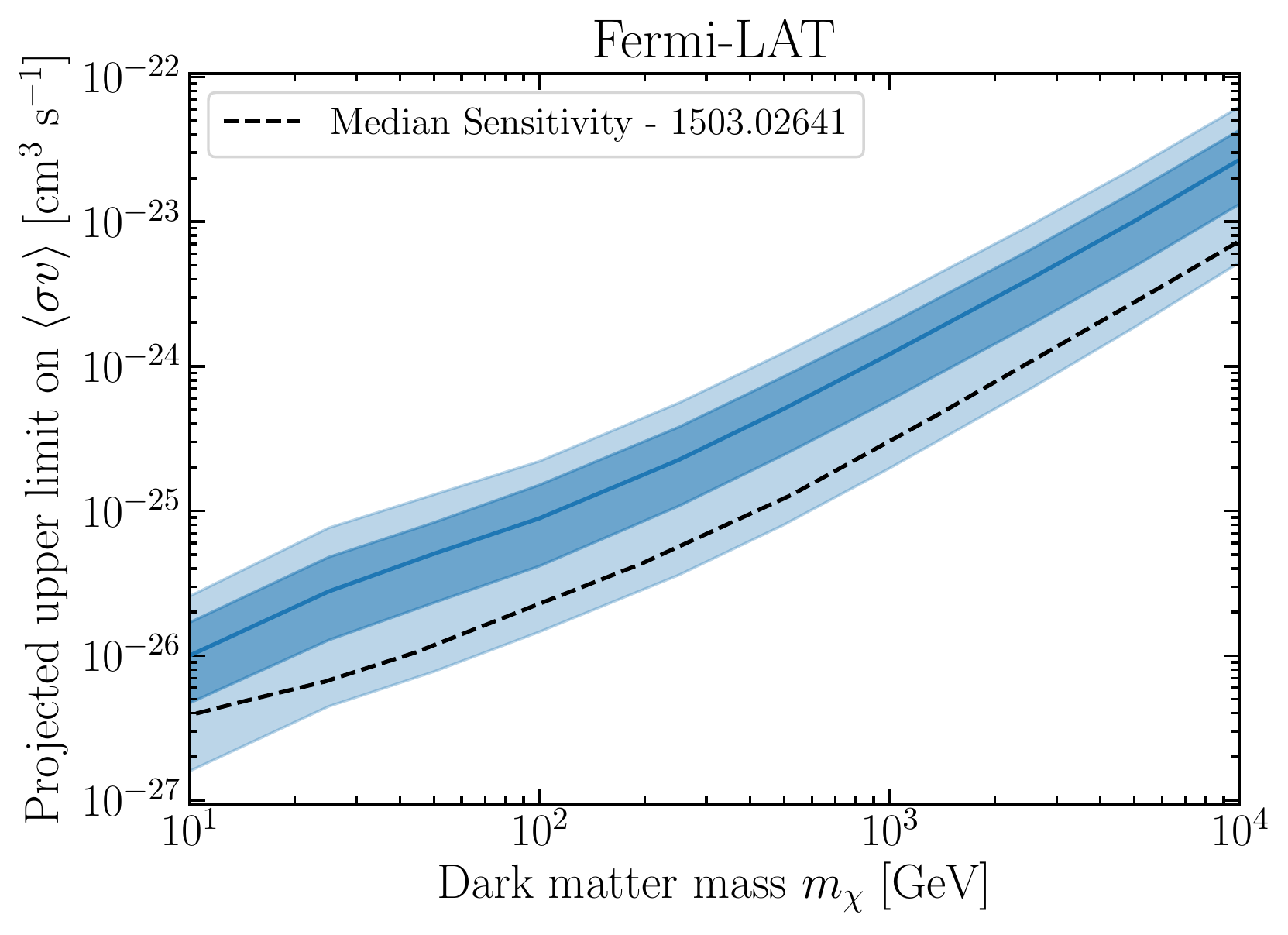}
\includegraphics[width=8.5cm]{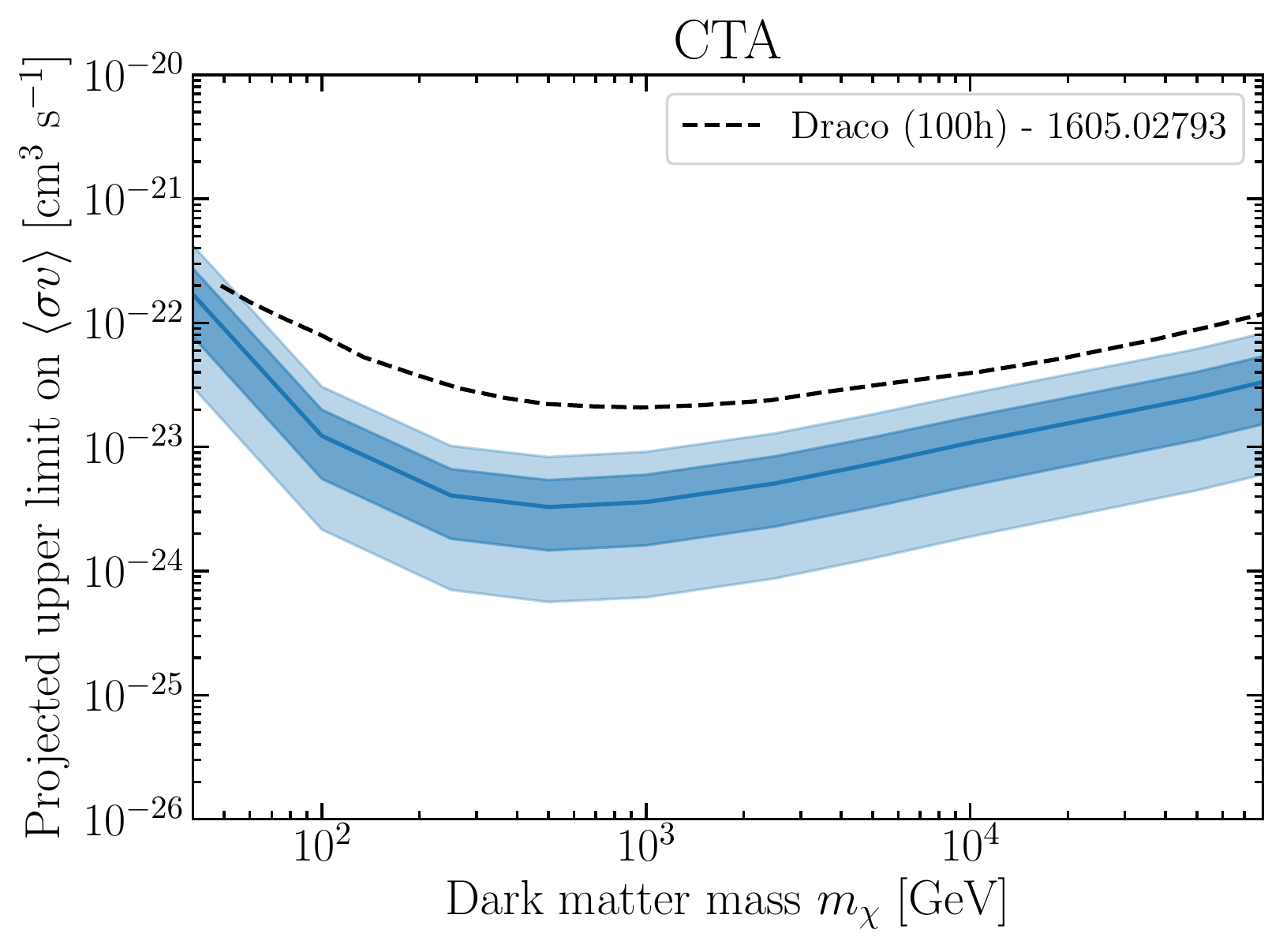}
\caption{Projected Fermi-LAT (top) and CTA (bottom) sensitivities for the dark matter annihilation cross section at 95\% confidence level, by using dSphs that can be found with LSST. The median and 68\% and 95\% containment regions are shown as a solid curve and filled bands. For comparison, we show median sensitivities by using existing dSphs for Fermi-LAT~\cite{Ackermann:2015zua} and Draco for CTA~\cite{Lefranc:2016dgx} as dashed curves.}
\label{fig:UpLimMass}
\end{figure}

We shall first study how the upper limits on the dark matter
annihilation cross section improve when more dSphs are included in
the analysis.
The variation of the expected upper limits on the annihilation cross
section as a function of the number of dSphs (sorted following $J$
factors from large to small values) is shown in Fig.~\ref{fig:CSJFac} for ten Monte Carlo realizations. Here, we assume observations with Fermi-LAT and a dark matter mass of
100~GeV.
One can see in Fig.~\ref{fig:CSJFac} that the sensitivity to the
annihilation cross section rapidly saturates after using a few to at most $\sim$10 dSphs
with the highest $J$-factors.

Figure~\ref{fig:UpLimMass} shows the
projected sensitivities to the annihilation cross section at 95\% 
confidence level for the Fermi-LAT (upper panel) and CTA (lower panel).
The bands contain 68\% and 95\% of the expected sensitivities with the
median shown as a curve in the middle.
These bands were constructed by performing 200 Monte-Carlo
simulations of the realizations of the Milky-Way dSphs that were placed
randomly within the LSST footprint, which causes a different background
for each dSph.
We note, however, that the {\sf swordfish} package provides only the \textit{median} expected 
sensitivity to a signal from a given realisation of dSphs. This means that the results do {\it not} include the uncertainties
associated with Poisson noise that arises from counting gamma-ray
photons.  According to Ref.~\cite{Fermi-LAT:2016uux} based on 6-year's Fermi-LAT
data, this photon noise further brings uncertainties up to one order of magnitude, which is
comparable to the bands associated with realization of dSphs shown in
Fig.~\ref{fig:UpLimMass}.

Our expected sensitivities with LSST dSphs will be able to exclude the canonical annihilation cross section of $\sim 2\times 10^{-26}$~cm$^3$~s$^{-1}$ for dark matter masses of only $\sim$20 GeV or lighter, although favorable statistical fluctuations at the $2\sigma$-level can bring this to about one order of magnitude heavier. The sensitivities for CTA, on the other hand, are unlikely to probe the canonical cross section of $\sim 2\times 10^{-26}$~cm$^3$~s$^{-1}$ for dark matter with masses between 50~GeV and tens of TeV.
They can probe at most the $10^{-24}$~cm$^3$~s$^{-1}$ regime.
However, Sommerfeld enhancement might bring the annihilation cross section for wino dark matter with masses around $\sim$3~TeV much larger than $10^{-24}$~cm$^3$~s$^{-1}$~\cite{Catalan:2015cna}, enabling us to test these models with the LSST dSphs.

We also compare our results to the Fermi-LAT joint likelihood analysis of tens of existing dSphs~\cite{Ackermann:2015zua} (to which the current upper limits are very similar~\cite{Fermi-LAT:2016uux}) and to projected constraints from 100 hours of exposure of CTA toward Draco~\cite{Lefranc:2016dgx}.
Our median sensitivities are weaker than those of Ref.~\cite{Ackermann:2015zua} by a factor of several. Part of this difference comes from the difference in exposures (an effective exposure of 2 years for this study, compared with 6 years in Ref.~\cite{Ackermann:2015zua}), while another source of discrepancy may be  the fact that our projection is based on a relatively simplified analysis (e..g, of using single pixels around each dSph with the radius of 0.5$^\circ$). Also, as we saw in Figs.~\ref{fig:J_dist_muV} and \ref{fig:J_dist_Vpeak}, our model predicts fewer dSphs with very large $J$-factors compared to the observed Milky Way population, leading to weaker projections. Again, this may be attributed to a statistical fluctuation or to optimistic estimates of the $J$-factors in earlier studies.

Recently, a similar prediction has been made for indirect dark matter searches with LSST dSphs in Ref.~\cite{Drlica-Wagner:2019xan}, presenting slightly stronger projected limits than those shown here. Our approach is more conservative for the same reasons as those listed above. In addition, Ref.~\cite{Drlica-Wagner:2019xan} assume 18 operational years compared with $\sim$10 years in our analysis. They combine expected LSST dSphs with the existing ones to obtain projected limits, while our analyses are based on the newly discovered LSST dSphs alone. These factors may explain the differences from Ref.~\cite{Drlica-Wagner:2019xan}.

For CTA, on the other hand, we find that the analysis of the LSST dSphs will improve the sensitivities further by up to one order of magnitude, compared with the estimates for Draco alone~\cite{Lefranc:2016dgx}.
Our estimates here are not unreasonable as we assume 50 hours of exposure for each dSph.
But as we demonstrated in Fig.~\ref{fig:CSJFac}, the sensitivity quickly saturates after fewer than 10 dSphs.
Therefore, there is no need to observe all $\sim$100 dSphs with 50 hour exposures for each.

\section{Discussion and conclusions}
\label{sec:Discussion and conclusions}

In this work, we have developed a realistic model for the distribution of Milky Way dwarf spheroidals (dSphs), based on semi-analytic models for dark matter subhalos and the satellite galaxies which form within them. We have then studied the prospects for observing these dSphs with the upcoming telescope LSST and for using them to constrain self-annihilating dark matter with gamma-ray telescopes such as the Fermi-LAT and the upcoming CTA.

We find that LSST should be able to discover most of the Milky Way dSphs within its footprint over several years of operation. The number of these discovered dSphs with large $J$-factors (those most relevant for the indirect detection of dark matter) is largely insensitive to the detailed thresholds for satellite formation. However, there is a factor of a few uncertainty in the number of dSphs which will be discovered due to Poisson uncertainties (we observe only one realisation of the Milky Way). Even so, we expect that LSST will discover 2--3 times as many dSphs with $J > 10^{18}$~GeV$^2$~cm$^{-5}$ as are currently known.

We find that with 2 years of effective exposure on all to-be-discovered LSST dwarfs, Fermi-LAT will be able to probe down to a cross section of $\langle \sigma v \rangle \sim 10^{-26}$~cm$^3$~s$^{-1}$ (for dark matter particles of mass 10~GeV). Instead, the upcoming CTA with 50 hours on each dwarf should constrain cross sections of $5\times 10^{-24}$~cm$^3$~s$^{-1}$ (for dark matter particles of mass 500~GeV). However, Poisson uncertainties on the number and properties of the discovered dwarf population translates into an order of magnitude uncertainty in the projected constraints.

We find that the joint analysis of the to-be-discovered dSphs with CTA should provide stronger constraints (by around an order of magnitude) than current projections with single dSphs. For Fermi-LAT, however, we find that our projections with LSST lie above the current Fermi-LAT sensitivity (though within the Poisson uncertainties). This is because our model predicts on average fewer dSphs with very large $J$-factors. It may be that the current $J$-factor estimates are optimistic, or that the Milky Way simply has more than the average expected number of very large dwarfs. In any case, we find that even with a large number of dwarf spheriodal galaxies to be discovered by LSST, constraints from the Fermi-LAT are unlikely to improve by more than a factor of a few and constraints from CTA are highly unlikely to reach down to the thermal freeze-out cross section.

\acknowledgments

We thank Ethan Nadler and Andrew Pace for helpful discussions.
This work was supported partly by GRAPPA Institute at the University of
 Amsterdam (SA and BJK) and JSPS KAKENHI Grant Numbers JP17H04836 (SA and OM), JP18H04578, and JP18H04340 (SA).

This project has been carried out in the context of the ``ITFA 
Workshop'' course, which is part of the joint bachelor programme in Physics and
Astronomy of the University of Amsterdam and the Vrije Universiteit
Amsterdam, for bachelor students (TA, SB, SD, TG, JG, JK, TM, JL, EP, BvdL, SV, and LXPV), supervised by SA, BJK, and OM.

The actual work was done in four independent groups A--D during a four-week period of January 2019.
The group A (SB, TG, and SV) worked on modeling dwarf galaxies in $V$-band magnitude and surface brightness, having contributed to Secs.~\ref{sub:Subhalo-satellite connection} and \ref{sec:surface_brightness}
 and Figs.~\ref{mv_distribution} and \ref{surface_brightness_distribution}.
The group B (LXPV, EP, and BvdL) obtained the $J$-factor distribution for the dwarfs, having contributed to Secs.~\ref{sub:Astrophysical J factor} and \ref{sec:Jfactor_distribution} and Figs.~\ref{fig:J_dist_muV} and \ref{fig:J_dist_Vpeak}.
The group C (JK and TM) worked on the spatial distribution of the satellite galaxies, having contributed to Sec.~\ref{sub:Subhalo distribution} and provided numerical codes to convert radial to Galactocentric coordinates to the other groups.
The group D (TA, SD, JG, and JL) performed the Monte Carlo sampling of the dwarfs based on the $J$-factor distribution, as well as the sensitivity projection for Fermi-LAT and CTA, having contributed to Secs.~\ref{sec:Prospects for gamma-ray observation} A--C and Figs.~\ref{fig:CSJFac} and \ref{fig:UpLimMass}.
The entire project was coordinated by the supervisors in group X (SA, BJK, and OM), who collected all the numerical codes developed by the students and brought the figures and text to refinement after the course had been completed.


\bibliographystyle{JHEP}
\bibliography{ITFA}

\end{document}